\documentclass[prd, twocolumn, showpacs, floatfix, superscriptaddress]{revtex4}
\usepackage[utf8]{inputenc}
\pdfoutput=1
 
\usepackage{graphicx}
\usepackage{epsfig}
\usepackage{bm}
\usepackage{amssymb}
\usepackage{float}
\usepackage{amsmath}
\usepackage{dcolumn}
\usepackage{cancel}
\usepackage[colorlinks]{hyperref}
\usepackage[usenames, dvipsnames]{color}

\hypersetup{
     breaklinks=true, 
    pdfstartview={FitH},  
    colorlinks=true, 
    linkcolor=blue,  
    citecolor=red,  
    filecolor=magenta,  
    urlcolor=blue, 
    anchorcolor=green,  
    linktocpage=true
}

\def\doi{http://doi.org}

\begin{document}

\title{ 
Information measures for fermion localization in $f(T, B)$ gravity with 
non-minimal couplings}

\author{Allan R. P. Moreira\footnote{Corresponding author}}
\email{allan.moreira@fisica.ufc.br}
\affiliation{Research Center for Quantum Physics, Huzhou University, Huzhou, 
313000, P. R. China.}
\affiliation{Secretaria da Educaç\~{a}o do Cear\'{a} (SEDUC), Coordenadoria 
Regional de Desenvolvimento da Educaç\~{a}o (CREDE 9),  Horizonte, Cear\'{a}, 
62880-384, Brazil.}
\author{Shi-Hai Dong}
\email{dongsh2@yahoo.com}
\affiliation{Research Center for Quantum Physics, Huzhou University, Huzhou, 
313000, P. R. China.}
\affiliation{Centro de Investigaci\'{o}n en Computaci\'{o}n, Instituto 
Polit\'{e}cnico Nacional, UPALM, CDMX 07700, Mexico.}

\author{Emmanuel N. Saridakis}
\email{msaridak@noa.gr}
\affiliation{National Observatory of Athens, Lofos Nymfon, 11852 Athens, Greece}
\affiliation{CAS Key Laboratory for Researches in Galaxies and Cosmology, 
School 
of Astronomy and Space Science, University of Science and Technology of China, 
Hefei, Anhui 230026, China}
\affiliation{Departamento de Matem\'{a}ticas, Universidad Cat\'{o}lica del 
Norte, Avda. Angamos 0610, Casilla 1280 Antofagasta, Chile}

\begin{abstract}

We investigate the dynamics of fermion localization within the framework of 
$f(T, B)$ gravity featuring non-minimal couplings. Starting from the Dirac 
action for a spin-$1/2$ fermion in a five-dimensional spacetime governed by 
torsional $f(T, B)$ gravity, we derive the  Dirac equation and we explore its 
solutions under various non-minimal coupling functions. We examine two realistic 
forms of the torsional non-minimal coupling and reveal distinct behaviors that 
impact the localization of both massless and massive fermionic modes on the 
brane. Additionally, we employ probabilistic measurements, including Shannon 
entropy theory, Fisher information theory, and relative probability, to analyze 
the localization of these fermionic modes. The observed effects offer potential 
insights into probing torsional modifications.  
\end{abstract} 
\maketitle

\section{Introduction}

The investigation of extra-dimensional theories traces its origins back to the 
pioneering works of Kaluza \cite{Kaluza1921} and Klein \cite{Klein1926} in the 
early 20th century. Initially motivated by the unification of Einstein's general 
relativity (GR) and Maxwell's electromagnetism, these theories opened new 
directions in theoretical physics. Much later, the possibility of non-compact 
extra-dimensional theories \cite{Arkani-Hamed:1998jmv, rs, rs2, dsh} broadened 
these directions further. In particular, the  Randall-Sundrum model paved the 
way for braneworld scenarios, while the Gregory-Rubakov-Sibiryakov model 
\cite{Gregory:2000jc}, the universal extra dimension model 
\cite{Appelquist:2000nn}, the Dvali-Gabadadze-Porrati model \cite{Dvali:2000hr}, 
and others, explored these possibilities even further, including the 
construction of thick branes using scalar, vector, and spinor fields 
\cite{DeWolfe:1999cp, Gremm1999, Csakil, Gremm2000, Dzhunushaliev:2009, 
Dzhunushaliev:2010, Herrera-Aguilar:2009, Dzhunushaliev:2007, 
Gogberashvili:2003a, Gogberashvili:2003b, Goldberger1999, Bazeia2008, 
Geng:2015kvs, Dzhunushaliev:2011mm}.{Although the brane model is still very theoretical, there are many studies that guide the possible detection of these particles that inhabit the extra dimensions, one possibility would be indirect detection in the LHC itself \cite{Frank:2016vtv,Angelescu:2017jyj,Goncalves:2017soe,CMS:2018vjd,Crivellin:2022nms}.}

On the other hand, theoretical motivations such as the non-renormalizability of  
GR, as well as cosmological requirements like solving the cosmological constant 
problem or addressing observational tensions (e.g., the $H_0$ and the $\sigma_8$ 
tensions) \cite{Abdalla:2022yfr}, have led to extensive research in the 
direction of modified gravity \cite{CANTATA:2021ktz, Capozziello:2011et, 
Cai:2015emx}. In this framework, various extensions and modifications of GR have 
been constructed to address these issues by introducing a richer structure. The 
simplest approach involves extending the Einstein-Hilbert action, resulting in 
$f(R)$ gravity \cite{Starobinsky:1980te, Capozziello:2002rd, DeFelice:2010aj}, 
$f(G)$ gravity \cite{Nojiri:2005jg, DeFelice:2008wz}, cubic gravity 
\cite{Asimakis:2022mbe}, Lovelock gravity \cite{Lovelock:1971yv, 
Deruelle:1989fj}, or scalar-tensor theories such as Horndeski gravity 
\cite{Horndeski:1974wa} and generalized Galileon theory \cite{DeFelice:2010nf, 
Deffayet:2011gz, Kobayashi:2010cm, DeFelice:2011bh}. Alternatively, one can 
start with the Teleparallel Equivalent of General Relativity (TEGR), where 
gravity is described by the torsion tensor, and extend it in various ways. This 
includes $f(T)$ gravity \cite{Cai:2015emx, Linder:2010py, Chen:2010va}, the 
Teleparallel Equivalent of Gauss-Bonnet and $f(T, T_G)$ gravity 
\cite{Kofinas:2014owa, Kofinas:2014daa}, and $f(T, B)$ gravity, where $B$ is the 
boundary term connecting the torsion and curvature scalars 
\cite{Bahamonde:2015zma, Bahamonde:2016grb}. Inspired by scalar-tensor theories 
\cite{Geng:2011aj}, one can also construct scalar-torsion theories by 
introducing a new scalar field coupled to torsion terms, resulting in the 
Teleparallel Equivalent of Horndeski gravity \cite{Bahamonde:2019shr, 
Bahamonde:2020cfv, Capozziello:2023foy, Aldrovandi}. Torsional modified 
gravities, particularly $f(T, B)$ theories, have been shown to have interesting 
cosmological and black hole phenomenology, thus attracting significant research 
interest \cite{Franco2020, EscamillaRivera2019, Bahamonde2016a, Abedi2017, 
Farrugia:2018gyz, Soudi:2018dhv, Gakis:2019rdd, Caruana2020, Pourbagher2020, 
Bahamonde2020a, Azhar2020, Bahamonde:2021dqn, Bhattacharjee2020, 
Mylova:2022ljr}.

Due to the success of braneworld constructions and modified theories of gravity, 
 a new direction has emerged: constructing braneworld scenarios within the 
framework of modified gravity. Hence,  the question of the 
location of matter fields in such modified braneworld scenarios arises 
naturally. The study of the location of fermions in teleparallel gravity models 
is rich and interesting, and can be addressed by considering fermionic 
Yukawa couplings in $f(T)$ gravity \cite{Yang2012} or $f(T, B)$ gravity 
\cite{Moreira20211}. However, although the Yukawa-type coupling is the most 
commonly used, it may not sufficiently show the influence of gravitational 
changes on the fermion's location in the brane. A potential solution is to 
propose new non-minimal coupling mechanisms between the fermion and the 
geometrical structure.

In this work, we investigate fermion localization in $f(T, B)$ gravity, allowing 
for  the possibility of non-minimal couplings, which are known to play an 
important role in teleparallel Horndeski theories. The structure of this paper 
is as follows: In Section \ref{sec1}, we study the influence of geometry on the 
location of fermions through a non-minimal coupling between the spinor field and 
torsional geometry. In Section \ref{sec2}, we apply probabilistic measurements 
as tools to further analyze how geometry affects the localization of the 
fermionic field on the brane. Finally, additional conclusions and comments are 
presented in Section \ref{finalremarks}.

\section{Fermion localization  in $f(T, B)$ gravity}
\label{sec1}

In this section, we investigate fermion localization within the framework of 
$f(T, B)$  gravity, incorporating non-minimal couplings between the fermionic 
field and the geometry. To begin, we briefly review the fundamentals of $f(T, 
B)$ gravity, working in $D$ dimensions for generality. Throughout this work, we 
set $\hbar=1$.

\subsection{Modified teleparallel gravity}
 
In torsional gravity, the dynamical fields are the vielbeins, which are related 
to the metric  through the following relationship:
\begin{eqnarray}
g_{MN}=\eta_{ab}h^a\ _M h^b\ _N, 
\end{eqnarray}
with $\eta_{ab}$ a $D$-dimensional Minkowski metric, and where capital Latin 
indices $M={0, ..., D-1}$ denote the coordinates in the bulk, while small Latin 
indices $a={0, ..., D-1}$ {denote the tangent space}. Furthermore, one 
introduces the Weitzenb\"{o}ck connection 
  \cite{Aldrovandi}
\begin{eqnarray}\label{a.685}
\widetilde{\Gamma}^P\ _{NM}=h_a\ ^P\partial_M h^a\ _N, 
\end{eqnarray}
which corresponds to zero curvature,  and therefore the corresponding torsion tensor is   
\begin{eqnarray}
T^{P}\ _{MN}= \widetilde{\Gamma}^P\ _{NM}-\widetilde{\Gamma}^P\ _{MN}, 
\end{eqnarray}
which quantifies all information about the gravitational field. 
The relationship between the Weitzenb\"{o}ck and the Levi-Civita 
connections is     $\widetilde{\Gamma}^P\ _{NM}= \Gamma^P\ _{NM} + K^P\ 
_{NM}$, where
\begin{eqnarray}
K^P\ _{NM}=\frac{1}{2}\Big( T_N\ ^P\ _M +T_M\ ^P\ _N - T^P\ _{NM}\Big), 
\end{eqnarray}
is the contortion tensor. Furthermore, it proves convenient to define the tensor
\begin{eqnarray}
S_{P}\ ^{MN}=\frac{1}{2}\Big( K^{MN}\ _{P}-\delta^N_P T^{QM}\ _Q+\delta^M_P 
T^{QN}\ _Q\Big).
\end{eqnarray}
Finally, we can define the torsion scalar as
\cite{Aldrovandi}
\begin{eqnarray}
T=T_{PMN}S^{PMN}.
\end{eqnarray}
If one uses $T$ as a Lagrangian, namely consider the action
\begin{equation}
    S=\int d^{D} x \, h\,  T, 
\end{equation} 
where $h=\rm det(h^{A}{}_{M})$ is the vielbein determinant.   By performing 
variation with respect to the vielbein, one obtains exactly the same equations 
as in GR, and this equivalence is why the theory in four dimensions is named 
TEGR. 
The underlying reason for this equivalence is that the torsion scalar 
corresponding to the Weitzenb\"{o}ck connection is related to the Ricci scalar 
$R$ corresponding to the Levi-Civita connection through
\begin{eqnarray}
-T+B=R, 
\end{eqnarray}
where
\begin{eqnarray}
B=-2\nabla^{M}T^{N}\ _{MN}, 
\end{eqnarray}
is a boundary term. However, considering  functions of the above quantities does 
not lead to equivalent theories anymore, because functions of total derivatives 
are not total derivatives themselves. Therefore, a more general theory is 
$f(T,B)$ gravity, which is characterized by the action
\begin{equation}
    S=\int d^{D} x\, h\, f(T, B).
    \label{action11}
\end{equation}
When $f(T, B)=f(T)$, the theory above corresponds to $f(T)$ gravity.  When $f(T, 
B)=f(-T+B)$, it corresponds to $f(R)$ gravity. Finally, when $f(T, B)=-T+\rm 
const$, the theory reproduces the TEGR, which includes GR along with a 
cosmological constant.

\subsection{Braneworld scenario and fermion localization}

Let us now proceed to the fermion localization in a 5-dimensional braneworld
scenario embedded in a bulk governed by $f(T, B)$ 
gravity. We consider  the metric ansatz that 
describes such a braneworld scenario as follows
\begin{equation}\label{45.a}
ds^ 2=e^{2A}(\eta ^{\mu\nu}dx^\mu dx^\nu+dz^2), 
\end{equation}
where $\eta_{\mu\nu}=(-1, 1, 1, 1)$ represents the four-dimensional  Minkowski 
metric and $e^{A(r)}$ is the warp factor that controls the thickness of the 
brane.
{In this scenario the brane ($4D$ 
Minkowski space)  is submerged in the bulk ($5D$ space).}
Specifically, we consider the warp factor in the form obtained in
\cite{Gremm1999}, namely
{
\begin{eqnarray}\label{20}
A(z)=-p\ln[\cosh(\mathrm{arcsinh}(\lambda z))], 
\end{eqnarray}}
where $p$ and $\lambda$ are parameters that modify the warp variation and 
determine the width within the brane core, respectively.

Now, apart from the bulk $f(T, B)$ gravity action (\ref{action11}), we consider 
that  the fermionic matter field $\Psi$ with spin $1/2$  is coupled to the 
torsional geometry in a non-minimal way. In particular, we consider the action 
\begin{eqnarray}\label{1}
\mathcal{S}_{1/2}=\int h \overline{\Psi} \Big(\Gamma^M D_M\Psi -\xi 
f(T, B)\Psi\Big)d^5x, 
\end{eqnarray}
where we use the gravitational model $f(T, B)$ as a non-minimal coupling function.
Here, $\Gamma^M=h_{a}\ ^M \Gamma^{a}$ represents the curved Dirac matrices
($\Gamma^{a}$ being the flat Dirac matrices) and $ D_M=\partial_M +\Omega_M$ 
denotes  the covariant derivative. Moreover, the spin connection $\Omega_M$ is 
expressed as \cite{Obukhov2002, Ulhoa2016, Moreira20211}
\begin{eqnarray}\label{3}
\Omega_M=\frac{1}{4}\Big(K_M\ ^{bc}\Big)\ \Gamma_{b}\Gamma_{c}.
\end{eqnarray}
We mention here that in (\ref{1}) one could have a more general 
coupling function $g(T,B)$ that is different from the gravitational sector 
$f(T,B)$, or even   consider that the gravitational 
sector is standard gravity, namely the teleparallel equivalent of general 
relativity (where the Lagrangian is just $-T$), and use the term
$f(T,B)$ only for the nonminimal coupling. The only reason that we use 
the same function is in order for our model to be \textcolor{blue}{simplified} (note that this 
consideration is used in other gravitational mechanisms too, for instance in 
gravitational baryogenesis, where   one 
 uses the same modified gravity term for the underlying gravitational 
evolution and the gravitational mechanism  \cite{Oikonomou:2016jjh}). In any 
case, as we will see below, our results qualitatively arise from the coupling 
function 
not from the underlying gravitational sector.

For our specific scenario, the spinor representation is given by 
\cite{Almeida2009, Dantas}
\begin{eqnarray}
\Psi\equiv\Psi(x, z)=\left(\begin{array}{cccccc}
\psi\\
0\
\end{array}\right)\ \mathrm{and}\ \
\Gamma^{a}=\left(\begin{array}{cccccc}
0&\gamma^{a}\\
\gamma^{a}&0\
\end{array}\right).
\end{eqnarray}
Consequently, the Dirac equation takes the form:
\begin{eqnarray}\label{7}
\Big[\gamma^{\mu}\partial_\mu+\gamma^4\partial_z-\xi e^Af(T, B)\Big]\psi=0.
\end{eqnarray}
Decomposing Eq.(\ref{7}) using 
$\psi=\sum_n[\psi_{L, n}(x)\varphi_{L, n}(z)+\psi_{R, n}(x)\varphi_{R, n}(z)]$, 
where $\gamma^4\psi_{R, L}=\pm\psi_{R, L}$ and 
$\gamma^\mu\partial_\mu\psi_{R, L}=m\psi_{L, R}$, yields two coupled equations, 
namely
\begin{eqnarray}\label{9}
\Big[\partial_z+\xi e^A f(T, B)\Big]\varphi_{L}(z)=m \varphi_{R}(z), \nonumber\\
\Big[\partial_z-\xi e^A f(T, B)\Big]\varphi_{R}(z)=-m \varphi_{L}(z).
\end{eqnarray}
These equations can be decoupled, leading to Schr\"{o}dinger-like 
equations
\begin{eqnarray}\label{10}
\Big[-\partial^2_z+V_L(z)\Big]\varphi_{L}(z)=m^2 \varphi_{L}(z), \nonumber\\
\Big[-\partial^2_z+V_R(z)\Big]\varphi_{R}(z)=m^2 \varphi_{R}(z), 
\end{eqnarray}
with the potential functions $V_L(z)$ and $V_R(z)$ defined as
\begin{eqnarray}\label{11}
V_L(z)=U^2 -\partial_{z}U, \nonumber\\
V_R(z)=U^2 +\partial_{z}U, 
\end{eqnarray}
and $U=\xi e^A f(T, B)$ representing the superpotential.

{It is interesting to highlight that Eqs.(\ref{10}) can be rewritten as:\begin{eqnarray}\label{10.1}
H_L\varphi_{L}(z)=m^2 \varphi_{L}(z), \nonumber\\
H_R\varphi_{R}(z)=m^2 \varphi_{R}(z), 
\end{eqnarray}
where the corresponding Hamiltonians are
\begin{eqnarray}\label{10.2}
H_L=A^{\dagger}A\ \ \mathrm{and}\ \
H_R=AA^{\dagger}, 
\end{eqnarray}
where $A^{\dagger}=-\partial_z+U$ and $A=\partial_z+U$. This shows that $H_L$ and $H_R$ are conjugate Hamiltonians of supersymmetric quantum mechanics and $V_L(z)$ and $V_R(z)$ are potential superpartners. The structure of Hamiltonians guarantees that the absence of tachyonic modes is real and that there are well-localized zero modes \cite{Almeida2009,Cooper:1994eh}.}

In order to ensure the localization on the brane and  self-consistency of our considerations, certain conditions on the form of $f(T, B)$ must be imposed. Drawing inspiration from the Yukawa-type coupling, which is prevalent in the literature, we propose the following conditions for $f(T, B)$:
\begin{enumerate}
    \item The function $f(T(z), B(z))$ should exhibit an isometric behavior, 
undergoing a phase transition at the origin ($z=0$).
    \item The $f(T, B)|_{z\rightarrow\pm\infty}$ should approach a constant 
value 
$c$ asymptotically.
    \item For our model to retain physical validity, the so-called 
superpotential $U|_{z\rightarrow\pm\infty}$ must tend to zero in
vacuum.
\end{enumerate}

Obeying these conditions, we define two forms for the geometric coupling 
function, namely, 
\begin{eqnarray}
 f_1(T, B)=\sqrt{-T-\alpha(B)}
 \label{g1eq}
\end{eqnarray}
and
\begin{eqnarray}
 f_2(T, B)=\sqrt{-T}-\alpha B, 
  \label{g2eq}
\end{eqnarray}
{where the parameter $\alpha$ controls the effect of the 
boundary term on the model.}

\begin{figure} []
\includegraphics[scale=0.68]{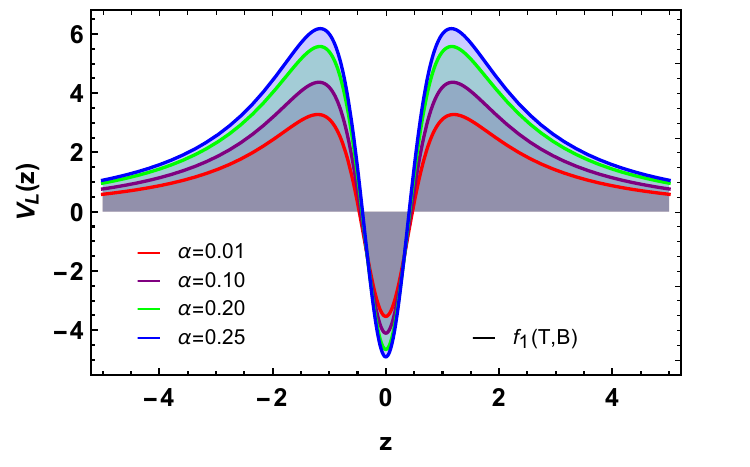} \\
\includegraphics[scale=0.68]{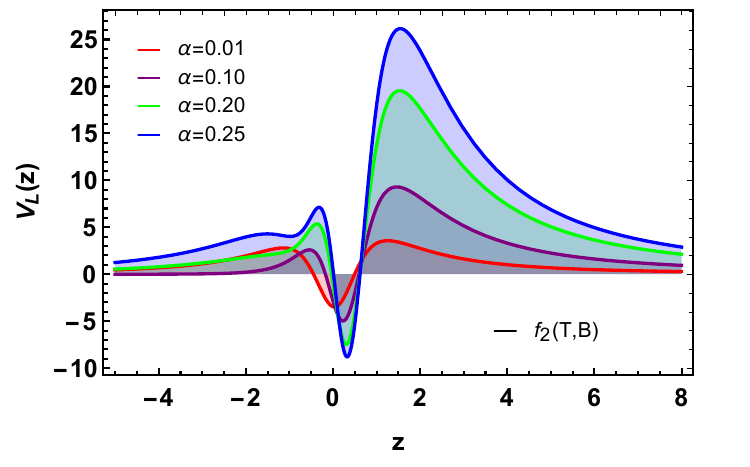}
\caption{ {{The effective potential for $\lambda=p=\xi=1$, and various 
values of the modified gravity parameter $\alpha$, for two cases 
of the $f(T, B)$ function, i.e.,   $f_1$ of (\ref{g1eq}), and 
$f_2$ of (\ref{g2eq}).}}
}
\label{fig1}
\end{figure}

For the first  choice of $f_1$, the effective potential (\ref{11}) exhibits a 
standard behavior commonly discussed in the literature, depicting a perfectly 
symmetric brane (see $f_1$ of Fig. \ref{fig1}). This outcome is significant 
because it ensures the localization of fermionic modes, thereby validating the 
physical viability of our model. Moving forward, we observe how variations in 
the geometry of the model influence the potential's shape. Increasing the 
parameter $\alpha$ enhances both the potential barriers and the depth of the 
potential well. This feature suggests that a greater contribution from the 
boundary term in our model enhances the localization of fermionic modes on the 
brane.

Our second choice, $f_2$,  represents a novel aspect of this work, illustrating 
how the underlying geometry induces defects in brane symmetry. As depicted in 
$f_2$ of Fig. \ref{fig1}, when $\alpha\rightarrow0$, the potential exhibits a 
perfect and symmetric solution. However, as the influence of the boundary term 
on the model increases, the potential solution deforms, showing a slight 
asymmetry. Importantly, the likelihood of localizing fermionic modes is higher 
on the right side of the brane. This asymmetry signifies a distorted brane 
solution, demonstrating that model geometry can indeed induce deformations on 
the brane. Notably, similar results have been observed in other studies 
\cite{Battisti:2008am, Perez-Payan:2011cvf, Sabido:2015xfk, Lopez:2017xaz, 
Lopez-Picon:2023rhc}.

This asymmetric behavior arises from the chosen function $f_2(T,B)$ itself. As 
the value of $\alpha$ increases, it tends to move completely towards $z>0$, 
 namely the boundary term causes an asymmetry in the function $f_2$, which 
leads   to an asymmetric brane structure whose solutions tend to move towards 
positive values of $z$. This reveals an interesting behavior, namely that with a
  purely geometric coupling we can create asymmetric brane structures. This 
feature may have  important implications, since in the study of cosmic 
evolution in a braneworld scenario built on an asymmetric configuration, 
asymmetric deformation contributes to accelerating   Universe 
\cite{Lima:2024qsv,Zhong:2023eeq,Rosa:2022fhl}.

In the following subsections we investigate the massless and massive  fermionic 
modes separately.

\subsubsection{Massless modes}

Similarly to the Yukawa-type coupling scenario, the specific form of the 
non-minimal coupling $f(T, B)$ plays a crucial role in governing the dynamics of the Kaluza-Klein states. Massless modes, also known as zero modes, exhibit the following form
\begin{eqnarray}
\varphi_{R0, L0}(z)\propto \exp{\Bigg[\pm\int\xi f(T,B)e^{A}dz\Bigg]}, 
\end{eqnarray}
owing to the supersymmetric structure of the potentials \eqref{11}.

To ascertain whether the zero modes can indeed be localized on the brane, we 
need to verify whether the normalization condition for the zero modes is 
satisfied, namely  whether we have that
\begin{eqnarray}
\int \varphi_{R0, L0}(z)^2dz<\infty.
\end{eqnarray}
Given that $f(T, B)e^{A}|_{z\rightarrow\pm\infty}\rightarrow0$, only the  
{left-chiral fermions ($\varphi_{L0}$)} are confined to the 
brane (for positive $\xi$), a 
feature shared by Yukawa coupling models \cite{Yang2012}. Therefore, 
defining $p=1$, we can derive a straightforward expression for the warp factor in conformal coordinates, namely $A(z)=-\ln\sqrt{1+\lambda^2z^2}$. As $z$ tends to infinity, $e^A\rightarrow1/\lambda|z|$, consequently, one has
\begin{eqnarray}
\varphi_{L0}(z\rightarrow\pm\infty)\rightarrow|z|^{-\frac{\xi}{\lambda} 
f(T, B)|_{\infty}}, 
\end{eqnarray}
where $f(T, B)|_{\infty}=c$. If the normalization condition holds, we obtain the following equivalent condition
\begin{eqnarray}
\int |z|^{-2\frac{\xi}{\lambda} f(T, B)|_{\infty}}dz<\infty.
\end{eqnarray}
The above integral converges only when $\xi>\lambda/2f(T, B)|_{\infty}$, 
implying that the {left-chiral fermions} can be localized on the brane under this condition.

\begin{figure}[]
\includegraphics[scale=0.68]{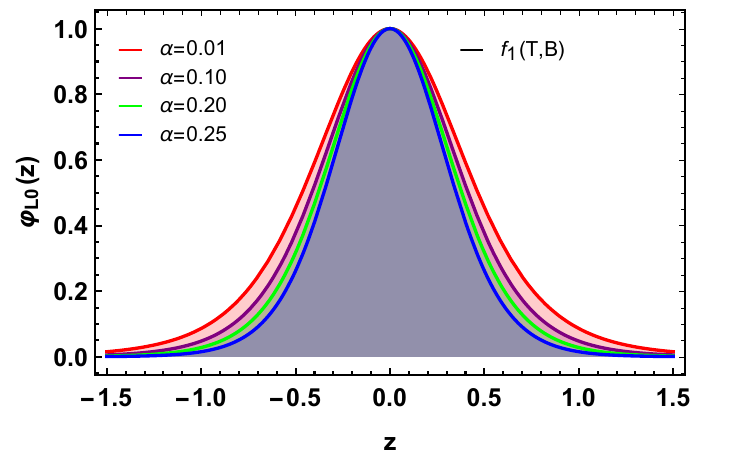} \\
\includegraphics[scale=0.68]{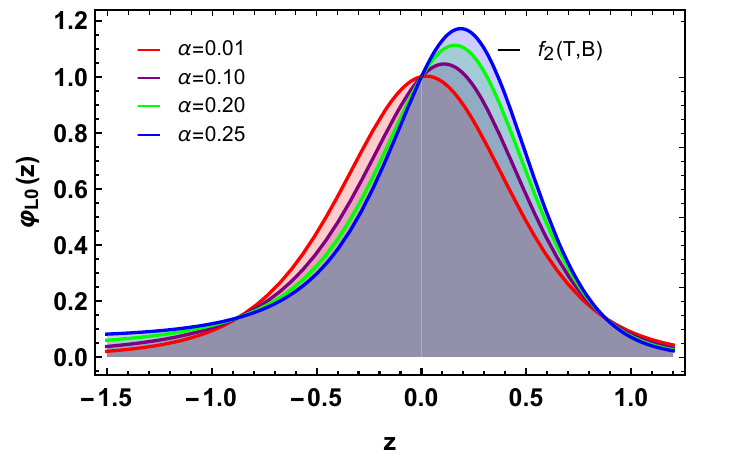}
\caption{ {{ The massless modes for $\lambda=p=\xi=1$, and various 
values of the modified gravity parameter $\alpha$,  for two cases 
of the $g(T, B)$ function, namely,  $g_1$ of (\ref{g1eq}), and 
$g_2$ of (\ref{g2eq}).}}
\label{fig2}}
\end{figure}

In Fig. \ref{fig2} we present the massless modes for the two choices of the 
$f(T, B)$ function. As we observe, the massless fermionic modes are well 
localized in our model. We mention that only the left-chirality fermions are 
localized in the brane. For $f_1$ case, the massless fermions feel the changes 
in the effective potential as $\alpha$ increases, and thus the massless modes 
become more localized in the brane (see $f_1$). Concerning the second case 
$f_2$, note that when $\alpha\rightarrow0$ the massless mode appears 
symmetrically at the origin of the brane. However, when we increase the effect 
of the boundary term, the zero-mode solutions become deformed and more localized 
on the right side of the brane (see $f_2$). This is an important feature, as 
these fermions, although localized, are located outside the core of the brane. 
This is one of the important results of the 
present work, since it may correspond to a possible observable of these 
fermionic particles and thus an indication for extra dimensions. 

\subsubsection{Massive  modes}

Let us now proceed to the determination of the massive fermionic modes.  To 
achieve this, considering the even symmetry of the effective potentials 
illustrated in Fig. \ref{fig1}, we will numerically solve equations (\ref{10}). 
It is crucial to note that the resulting wave functions will exhibit either even 
or odd symmetry. Consequently, the boundary conditions are set as follows
\begin{eqnarray}
\varphi_{even}(0)&=&c, \ \ \partial_z\varphi_{even}(0)=0, \nonumber\\
\varphi_{odd}(0)&=&0, \ \ \partial_z\varphi_{odd}(0)=c, 
\end{eqnarray}
where $c$ is a constant \cite{Almeida2009, Liu2009, Liu2009a}. Here, 
$\varphi_{even}$ and $\varphi_{odd}$  denote the even and odd parity 
modes of $\varphi_{R, L}(z)$, respectively.

\begin{figure} []
\includegraphics[scale=0.7]{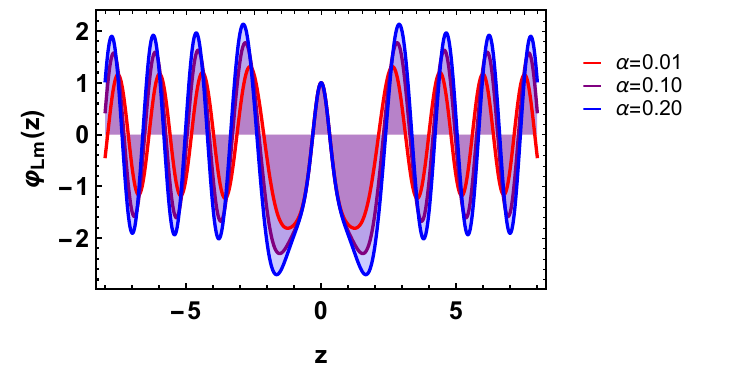} \\
\includegraphics[scale=0.7]{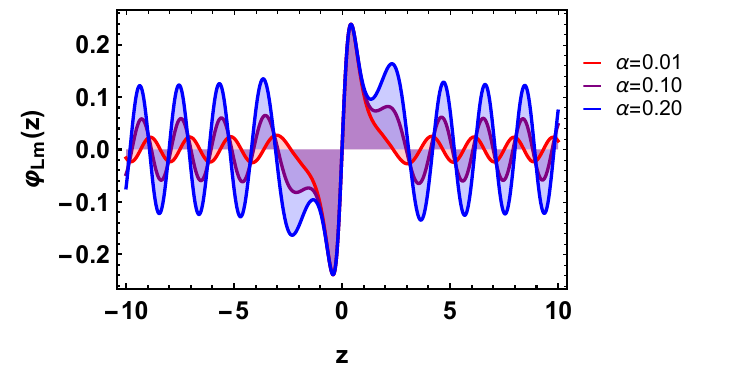}
 \caption{ {{
The massive modes for $\lambda=p=\xi=1$,  for the $f_1$ case of 
(\ref{g1eq}) and various 
values of the modified gravity parameter $\alpha$,  for even case (a), and odd case (b).}}
\label{fig3}}
\end{figure}

\begin{figure}[] 
\includegraphics[scale=0.7]{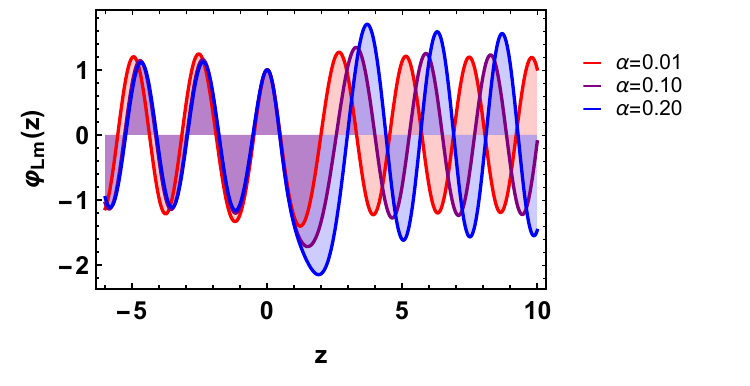} \\
\includegraphics[scale=0.7]{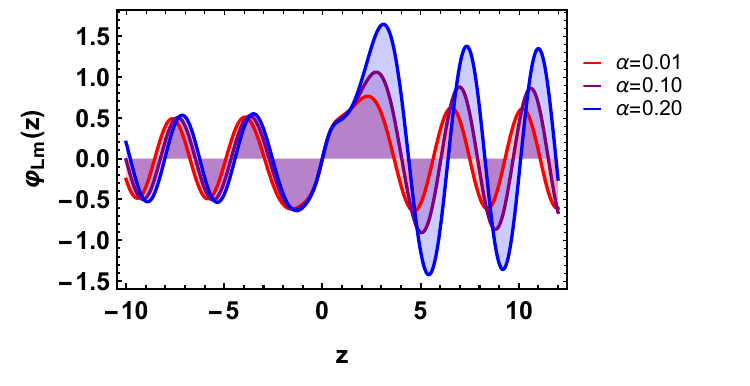} 
\caption{ 
{{ Same as Fig. \ref{fig3} but for the $f_2$ case of (\ref{g2eq})}}.
\label{fig4}}
\end{figure}

As observed, massive fermionic modes typically  resemble free wave solutions. 
This resemblance stems from the substantial energy that these massive fermions 
possess, ensuring that very massive fermions (high energy) tend to escape into 
the extra dimension. However, it is important to note that despite their 
tendency to escape into the bulk, these fermions can still sense geometric 
modifications near the brane core. In Fig. \ref{fig3}, we depict the solutions 
of the massive fermionic modes for the case $f_1$ of (\ref{g1eq}). It is evident 
that as the parameter $\alpha$ increases, the oscillations in the brane core 
undergo significant modifications. Thus, even though these modes are not 
localized on the brane, their influence can be detected there, potentially 
affecting measurements such as gravitational wave detections, leading to noise 
effects.

Moreover,  the massive fermionic modes for $f_2$ exhibit asymmetric behavior, as 
illustrated in Fig. \ref{fig4}. Increasing the influence of the boundary term in 
the model results in heightened oscillations on the right side of the brane. 
Noticeably, in this scenario, the brane behaves akin to a barrier, decreasing 
the likelihood of fermions crossing {the brane}.

\begin{figure*} 
\includegraphics[scale=0.45]{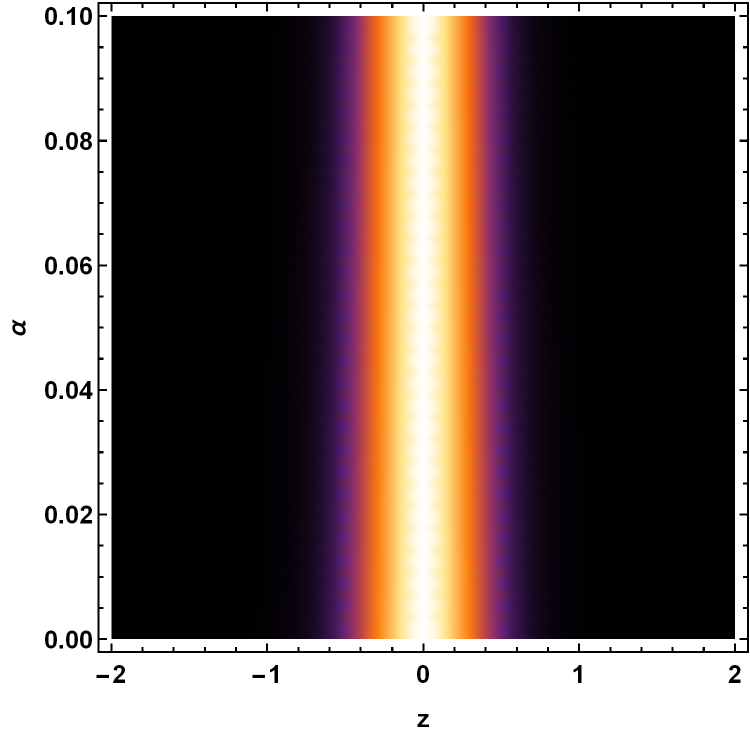} 
\includegraphics[scale=0.45]{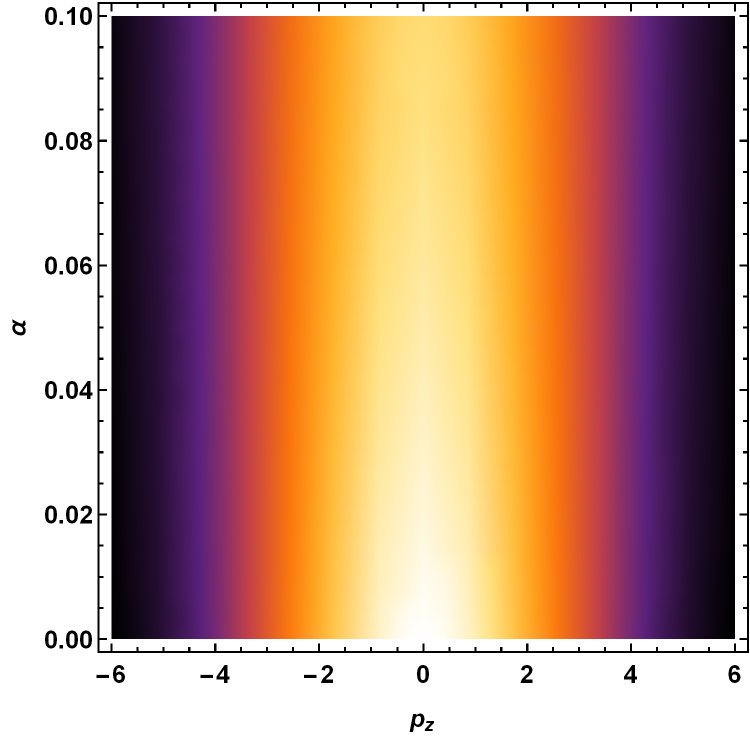}\\
\includegraphics[scale=0.45]{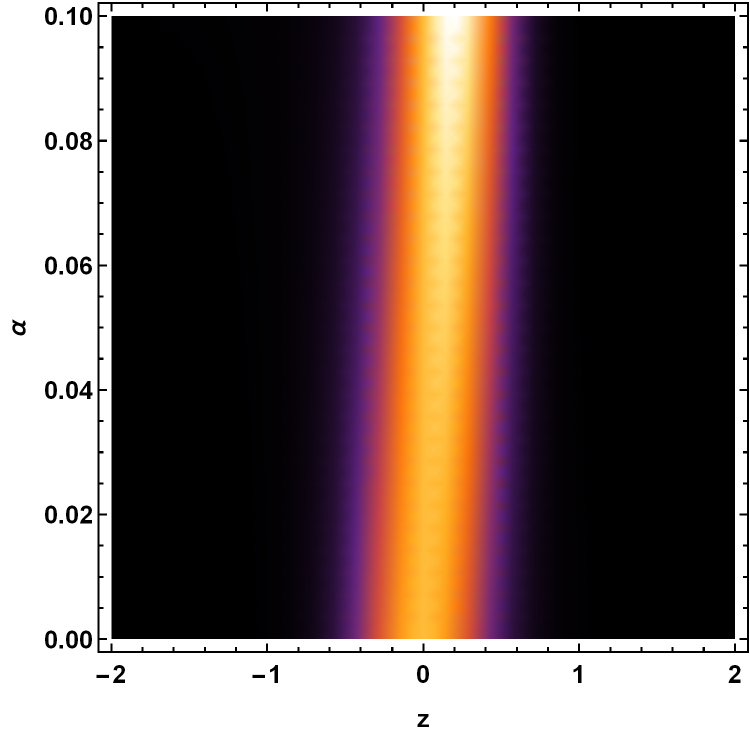} 
\includegraphics[scale=0.45]{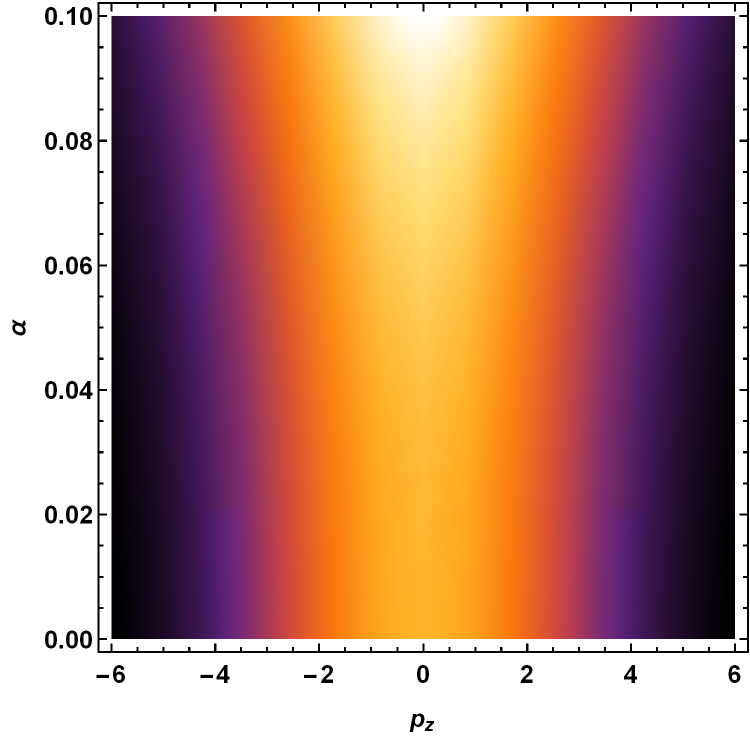}
\caption{{{ The entropy densities for $\lambda=p=\xi=1$. The left panels
  correspond to position space (\ref{0.11}), while the 
right panels, correspond to momentum space (\ref{0.11b}). Additionally, the 
upper panels,  correspond to the case $f_1$ of (\ref{g1eq}), while the lower 
panels correspond to the case $f_2$ of (\ref{g2eq}). Lighter regions correspond to higher probability of appearance than the darker 
ones.}}
\label{fig6}}
\end{figure*}

\section{Information measures}
\label{sec2}

{The motivation of the present study  is to uncover deeper 
connections between gravitational theories and quantum information measures, 
and that is why we extend traditional quantum information theory to 
the framework of $f(T, B)$ gravity with non-minimal couplings. Central to our 
analysis are the Shannon entropy and the Fisher information, which serve as 
essential tools for investigating the interplay between quantum uncertainty, 
information flow, and the geometry of spacetime in the context of modified 
gravity}.

{ In order to investigate these information measures, we 
calculate  the Shannon entropy \cite{Shannon} and the Fisher information 
\cite{Fisher}, focusing firstly on the massless case. For massive modes, 
however, the relative probability formalism is employed. Shannon entropy, a 
fundamental measure of uncertainty or information content, provides valuable 
insights into the distribution of quantum states in both position and momentum 
spaces. Meanwhile, Fisher information quantifies the sensitivity of a 
probability density function, $\rho(x) = |\psi(x)|^2$, to variations in the 
spatial coordinate $x$. In the context of $f(T, B)$ gravity, these measures 
enable a detailed characterization of how modifications in spacetime geometry 
affect quantum systems and their associated informational content}.

{By  calculating   Shannon entropy in both position and 
momentum spaces, we aim to reveal the effects of gravitational modifications on 
the localization of quantum states and their associated uncertainties. 
Simultaneously, Fisher information allows us to probe the sensitivity of quantum 
states to changes in spacetime parameters, shedding light on the effect of 
the gravitational framework on their informational structure}.

{Hence, our motivation and goals are the following:   
1) quantify the uncertainty in quantum states; 2) probe the sensitivity to 
gravitational parameters; 3) establish links between gravity and information, 
i.e. investigate the fundamental relation between the informational 
structure of quantum states and the geometric framework of modified gravity 
theories and 4) extend the benefit of quantum information measures to analyze 
quantum fields and their evolution in the $f(T,B)$ gravity framework}.

\subsection{Shannon theory}

Let us first examine the Shannon information entropy.
The Shannon entropy for both 
position and momentum spaces is defined as
\begin{eqnarray}\label{0.11}
S_{z}&=&-\int_{-\infty}^{\infty} \rho_s(z) dz, \\
S_{p_z}&=&-\int_{-\infty}^{\infty}\rho_s(p_z) dp_z, \label{0.11b}
\end{eqnarray}
where
\begin{eqnarray}\label{rhosz}
\rho_s(z)&=&\vert\varphi_{L0, R0}(z)\vert^{2}\ln\vert\varphi_{L0, 
R0}(z)\vert^{2}, 
\\
\rho_s(p_z)&=&\vert\varphi_{L0, R0}(p_z)\vert^{2}\ln\vert\varphi_{L0, R0}
(p_z)\vert^{2}  
\end{eqnarray}
are the entropy densities of the system.
We mention that the 
position Shannon entropy, given by   (\ref{0.11}), can be directly calculated 
from the position entropy density defined   in  (\ref{rhosz}). However, 
the calculation of the momentum Shannon entropy, given by   (\ref{0.11b}), is 
more intricate. This complexity arises  from the fact that it requires 
determining the wave function in momentum space, which involves performing a 
Fourier transformation of the massless mode function, expressed as
{
\begin{equation}\label{fouu}
\vert\varphi_{L0, R0}(p_z)\vert^{2}=\frac{1}{\sqrt{2\pi}}\int_{-\infty}^{\infty}
\vert\varphi_{L0, R0}(z)\vert^{2} \text{e}^{-ip_{z}} dz, 
\end{equation}}
where $p_z$ is the coordinate of momentum space or reciprocal space.

Relations (\ref{0.11}), (\ref{0.11b}) yield an uncertainty relation, 
known as the Beckner, Bialynicki-Birula, Mycielski (BBM) relation 
\cite{Beckner, Bialy}. The important aspect of this entropy 
uncertainty relation is its effectiveness as an alternative to the 
Heisenberg uncertainty principle. The BBM uncertainty relation is 
mathematically 
expressed as
\begin{equation}
S_{z}+S_{p_z}\geq D(1+\text{ln}\pi), 
\end{equation}
where $D$ denotes the dimensions capable of discerning changes in the
system information. In our model, only the extra dimension perceives entropy
modifications within the system, and therefore $D=1$.

\begin{figure*} 
\includegraphics[scale=0.45]{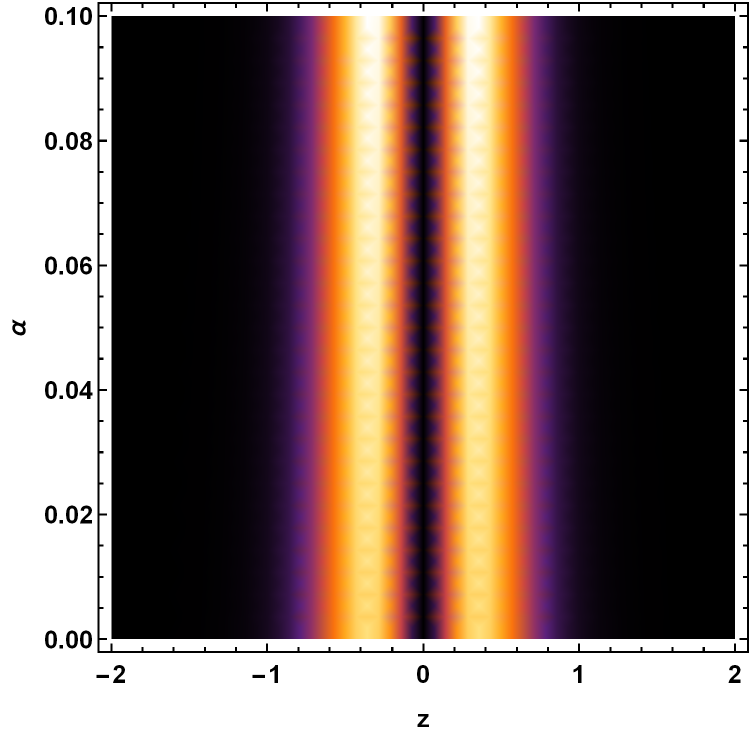} 
\includegraphics[scale=0.45]{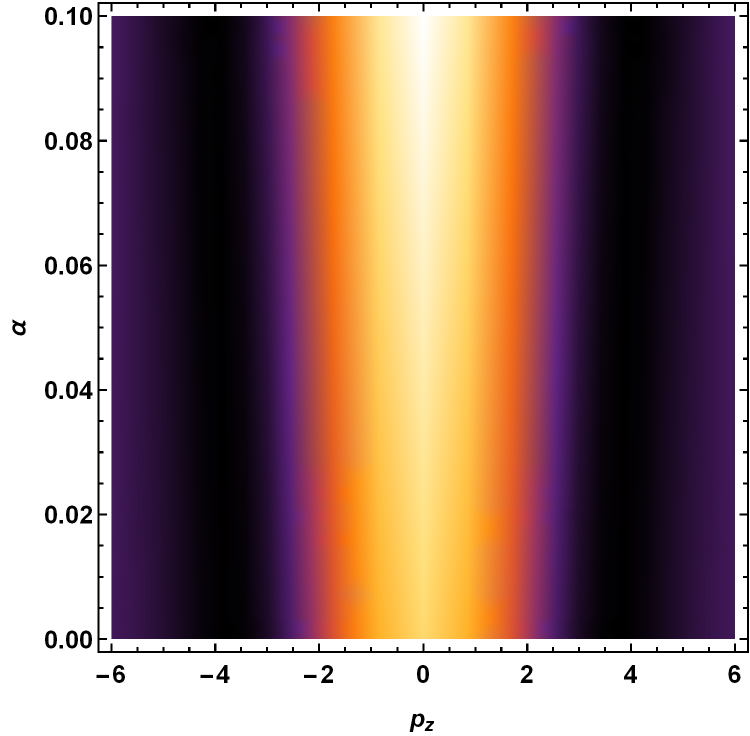}\\ 
\includegraphics[scale=0.45]{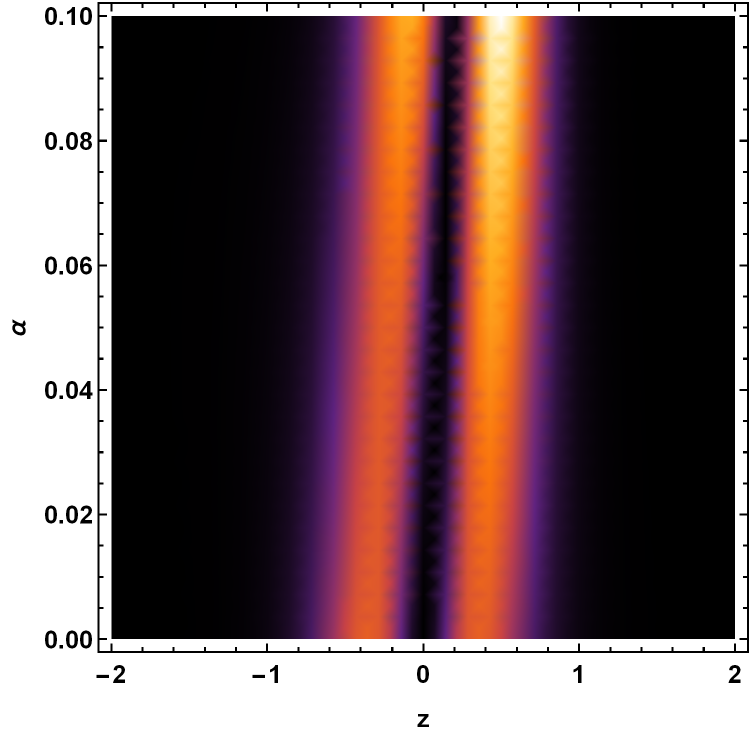} 
\includegraphics[scale=0.45]{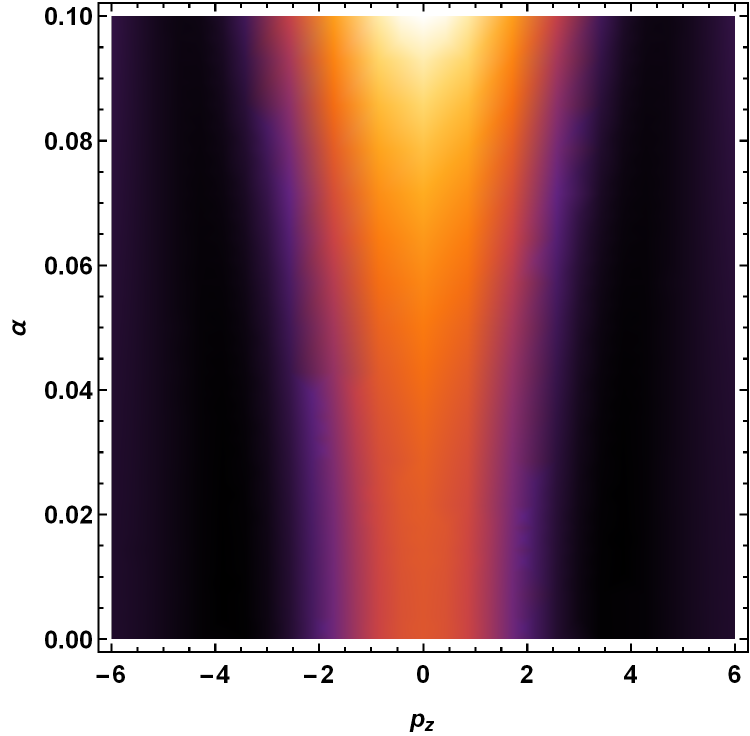} 
\caption{ 
{{ The information densities for $\lambda=p=\xi=1$. The left 
panels  correspond to position space (\ref{Fisher1}), while the 
right panels correspond to momentum space (\ref{Fisher2}). Additionally, the 
upper panels   correspond to the case $f_1$ of (\ref{g1eq}), while the lower 
panels   correspond to the case $f_2$ of (\ref{g2eq}).
Lighter regions correspond to higher probability of appearance than the darker 
ones.}}
\label{fig7}} 
\end{figure*}

In Fig. \ref{fig6}, we present the shape of the  entropy density for our two 
choices of the $f(T, B)$ modified gravity function of (\ref{g1eq}) and 
(\ref{g2eq}). Lighter regions correspond to a higher probability of appearance 
than darker ones. For the $f_1$ case, the entropy density in position space 
appears linearly at the origin of the brane, while in momentum space, the 
density appears thicker. In contrast, for the $f_2$ case, the entropy density 
deviates from the origin as we increase the value of $\alpha$.

{Finally, we numerically  compute the position and momentum 
Shannon entropy and we present the results in Table \ref{tab1} for the 
specified parameter values. As the effect of the boundary term $\alpha$ 
increases, the Shannon entropy in position space decreases, but the Shannon 
entropy in momentum space increases. This behavior is in agreement with the 
uncertainty principle and is consistent with the BBM relation. Furthermore, the 
total entropy increases with larger values of $\alpha$, reflecting a higher 
uncertainty of the system}.

\begin{table}  [ht]
\caption{Shannon entropy values in position and momentum space for the two cases of the $f(T, B)$ function, with $\xi=p=\lambda=1$.\label{tab1}}
\begin{tabular}{|c||c|c|c|c|}
\hline
$n$ & $\alpha$ & $S_{z}$ & $S_{p_z}$ & $S_{z}+S_{p_z}$\\ \hline
\hline
$g_1$  & 0.01  & 0.21998 & 2.28577  & 2.50575  \\
       & 0.10  & 0.20899 & 2.30903  & 2.51802  \\
       & 0.20  & 0.16477 & 2.40538  & 2.57015\\ 
       & 0.25  & 0.15473 & 2.42789  & 2.58262  \\ \hline \hline

$g_2$  & 0.01  & 0.22122 & 2.28318  & 2.50441   \\
       & 0.10  & 0.22023 & 2.28722  & 2.50745   \\
       & 0.20  & 0.19858 & 2.37979  & 2.57837 \\ 
       & 0.25  & 0.19099 & 2.41879  & 2.60978  \\ \hline \hline

\end{tabular}
\end{table}

\subsection{Fisher theory}

 While Shannon entropy focuses on the uncertainty about the 
system state,   Fisher information focuses on the distribution sensitivity   to 
changes in the parameters, and hence it is  relevant for parameter estimation.

Similar to Shannon entropy density, we can define the Fisher information density as follows 
\begin{eqnarray}
\label{Fisher1}
\rho_F(z)&=&\vert\varphi_{L0, R0}(z)\vert^{2}\bigg[\frac{d}{dz}\ln\vert\varphi_{
L0, R0}(z)\vert^2\bigg]^2, \\
\rho_F(p_z)&=&\vert\varphi_{L0, R0}(p_z)\vert^{2}\bigg[\frac{d}{dp_z}
\ln\vert\varphi_{L0, R0}(p_z)\vert^2\bigg]^2,
\label{Fisher2}
\end{eqnarray} 
from which we might calculate Fisher information by the following way
$F_{p_z}$, namely
\begin{eqnarray}
F_{z}&=&\int_{-\infty}^{\infty}\rho_F(z) dz, \label{f11}\\
F_{p_z}&=&\int_{-\infty}^{\infty}\rho_F(p_z) dp_z\label{f12}.
\end{eqnarray}
{It is worth noting that these probabilistic information 
measures     provide also an alternative formulation of the Heisenberg 
uncertainty principle, namely}:
\begin{equation}
\sigma_{z}\sigma_{p_z}\geq\frac{1}{(F_{z}F_{p_z})^{\frac{1}{2}}}\geq\frac{1}{2}, 
\end{equation}
leading to
\begin{equation}
F_{z}F_{p_z}\geq4, 
\end{equation}
where $\sigma_{z}^{2}=\langle z^{2}\rangle-\langle z\rangle^{2}$ and 
$\sigma_{p_z}^{2}=\langle p_z^{2}\rangle-\langle p_z\rangle^{2}$.

In Fig. \ref{fig7},  we present the behavior of the Fisher information density 
for the two choices of the $f(T, B)$ modified gravity functions, given by 
equations (\ref{g1eq}) and (\ref{g2eq}). In the $f_1$ case, we observe that 
regions of low Fisher information density are separated by regions of higher 
Fisher information density. Furthermore, in position space, there are two 
distinct peaks in the Fisher information density, whereas in momentum space, 
only one peak is observed. In the $f_2$ case, the Fisher information density 
highlights the deformation of the brane, exhibiting noticeable changes in both 
position and momentum spaces.

We numerically compute the Fisher  information and present the results in Table 
\ref{tab2}. As observed, increasing the influence of the boundary term enhances 
the Fisher information measures in position space. In contrast, in momentum 
space, the Fisher information measures decrease as $\alpha$ increases. 
{Thus, a higher Fisher information corresponds to lower 
uncertainty about the parameter being estimated.}

\begin{table}[ht]
\caption{Fisher information values in position and momentum space for the two cases of the $f(T, B)$ function, with $\xi=p=\lambda=1$. \label{tab2}}
\centering
\begin{tabular}{|c||c|c|c|c|}
\hline
$n$ & $k$ & $F_{z}$ & $F_{p_z}$ & $F_{z}F_{p_z}$\\ \hline
\hline
$g_1$  & 0.01  & 2.81545 & 7.74121 & 21.7949  \\
       & 0.10  & 2.87569 & 7.39004 & 21.2514  \\
       & 0.20  & 3.13204 & 6.02243 & 18.8623\\ 
       & 0.25  & 3.19351 & 5.72364 & 18.2785  \\ \hline \hline

$g_2$  & 0.01  & 2.80880 & 7.91084  & 22.2199   \\
       & 0.10  & 2.81871 & 7.22346  & 20.3608   \\
       & 0.20  & 3.05356 & 5.75099  & 17.5609 \\ 
       & 0.25  & 3.15738 & 5.29128  & 16.7065  \\ \hline \hline

\end{tabular}
\end{table}

\subsection{Relative probability}

Let us now focus on the massive fermionic modes. Although they do not reside on the brane, certain heavy states may exhibit significant presence in close proximity to it \cite{Almeida2009}. These substantial states manifest when the potentials near the brane form a potential well, accommodating masses $m^2$ within the highest value of the potential barrier \cite{Liu2009, Liu2009a}. We refer to these occurrences as resonant modes, which arise from the analogous quantum mechanical structure of heavy modes \cite{Almeida2009, Liu2009, Liu2009a}.
\begin{figure}[] 
\includegraphics[scale=0.7]{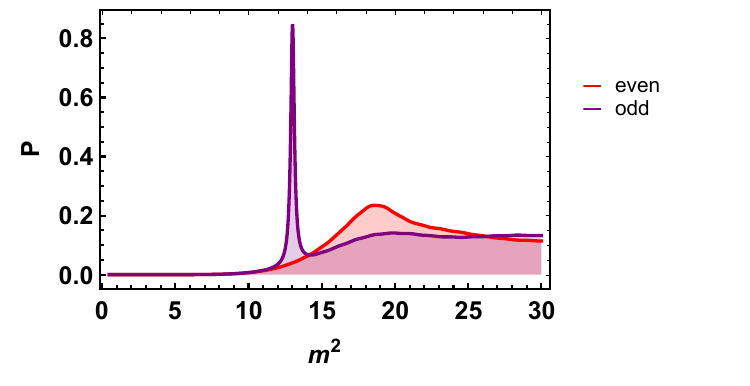}\\
\includegraphics[scale=0.7]{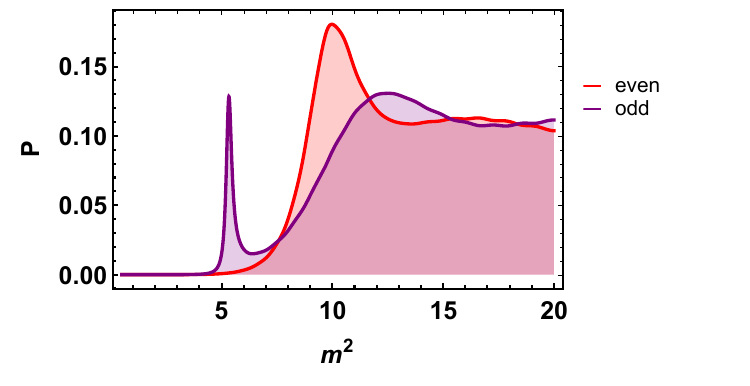}
\caption{{{ The relative probability (\ref{Prob1}) for $\xi=p=\lambda=1$ 
and $\alpha=0.1$, for two cases 
of the $g(T, B)$ function, namely:   $g_1$ of (\ref{g1eq}), and 
$g_2$ of (\ref{g2eq}).}}
\label{fig5}}
\end{figure}

To pinpoint the solutions of the Schr\"{o}dinger-like equation 
(\ref{10}) with significant amplitudes near the brane, we apply the resonance 
method. The relative probability $P(m)$ of detecting a particle with mass $m$ within a narrow band $2z_b$ is given by \cite{Almeida2009, Liu2009, Liu2009a}
\begin{eqnarray}
P_{R, L}(m)=\frac{\int_{-z_b}^{z_b} |\varphi_{R, L}(z)|^2 
dz}{\int_{-z_{\text{max}}}^{z_{\text{max}}} |\varphi_{R, L}(z)|^2 dz}, 
\label{Prob1}
\end{eqnarray}
where $z_{\text{max}}$ represents the domain boundaries. While larger values of the parameter $z_b$ maintain consistency in locating the resonance peaks, 
smaller $z_b$ values prove more adept at pinpointing these peaks.

We numerically compute the equations, and in Fig. \ref{fig5}, we illustrate the relative probabilities for our model. The peaks in relative density delineate resonant modes with the highest likelihood of locating massive fermions on the brane. For the $f_1$ case, the peak occurs around $12 < m^2 < 14$ (odd), while for the $f_2$ case, the peak is around $4.5 < m^2 < 6.0$ (odd). This indicates that in this model, massive fermions with these mass spectra have a greater chance of being detected on the brane.

\section{Conclusions}
\label{finalremarks}
In this work, we investigated the dynamics of fermion localization within the framework of $f(T, B)$ gravity. Starting from the Dirac action for a fermion with spin $1/2$ embedded in a 5-dimensional spacetime governed by torsional $f(T, B)$ gravity, we formulated the Dirac equation and explored its solutions under various non-minimal coupling functions.

We examined two realistic forms of the non-minimal coupling function $f(T, B)$, i.e., $f_1(T, B)=\sqrt{-T-\alpha(B)}$ and $f_2(T, B)=\sqrt{-T}-\alpha B$. Our analysis revealed distinct behaviors, highlighting implications for the localization of both massless and massive fermionic modes on the brane. Particularly, we observed how geometric deformations induced in the brane can influence the localization of fermionic modes, potentially providing avenues for experimental detection of these particles.

Furthermore, we employed probabilistic measures-Shannon entropy theory, Fisher information theory, and relative probability-to investigate the localization of massless and massive fermionic modes. Our analysis provided insights into the information content and uncertainty associated with fermion localization on the brane, offering a nuanced understanding of the interplay between position and momentum spaces.

In summary, this study represents a significant step toward comprehending the impact of modified gravity on fermion localization in brane world scenarios. The observed effects could potentially serve as probes for torsional modifications and may open new avenues for exploring holography and brane cosmography.

\section*{Acknowledgments}  SHD would like to thank the 
partial support of project 20240220-SIP-IPN, Mexico and started this work on the research stay in China.  
ENS acknowledges the contribution of the LISA CosWG, and of   COST 
Actions  CA18108  ``Quantum Gravity Phenomenology in the multi-messenger 
approach''  and  CA21136 ``Addressing observational tensions in cosmology with 
systematics and fundamental physics (CosmoVerse)''. 
{ARPM is grateful for the hospitality and support provided by the Research Center for Quantum Physics at Huzhou University. The authors also thank the anonymous referees for their valuable comments and suggestions.}

\end{document}